\documentclass[conference, 10pt]{IEEEtran}

\usepackage[cmex10]{amsmath}
\usepackage{amsfonts}
\usepackage{multirow}
\makeatletter
\usepackage{algorithm}
\usepackage{algorithmic}
\usepackage[tight,footnotesize]{subfigure}
\usepackage{graphicx}
\usepackage{color}
\usepackage{placeins}
\usepackage{float}
\usepackage{hyperref}
\usepackage{tabularx,colortbl}

\usepackage{lipsum}
\usepackage{mathtools}

\usepackage{cuted}

\usepackage{tikz}
\usepackage{textcomp}
\usepackage{hyperref}
\usepackage{lipsum}

\newcommand\copyrighttext{%
  \footnotesize \textcopyright 2017 IEEE. Personal use of this material is permitted.
  Permission from IEEE must be obtained for all other uses, in any current or future
  media, including reprinting/republishing this material for advertising or promotional
  purposes, creating new collective works, for resale or redistribution to servers or
  lists, or reuse of any copyrighted component of this work in other works.
}
\newcommand\copyrightnotice{%
\begin{tikzpicture}[remember picture,overlay]
\node[anchor=south,yshift=10pt] at (current page.south) {\fbox{\parbox{\dimexpr\textwidth-\fboxsep-\fboxrule\relax}{\copyrighttext}}};
\end{tikzpicture}%
}

\title{Direct Solution of FEM Models: Are  Sparse Direct Solvers the Best Strategy?}

\author{\IEEEauthorblockN{Javad Moshfegh and Marinos N. Vouvakis}
\IEEEauthorblockA{Department of Electrical and Computer Engineering, University of Massachusetts, Amherst, MA, USA}}
\begin{document}
\maketitle
\copyrightnotice

\begin{abstract}
A brief summary of direct solution approaches for finite element methods (FEM) in computational electromagnetics (CEM) is given along with an alternative direct solution based on  domain decomposition (DD). Unlike recent trends in approximate/low-rank solvers, this work focuses on `numerically exact' solution methods as they are more reliable for complex `real-life' models. Preliminary studies on general three dimensional geometries with unstructured FEM meshes suggest that the proposed direct DD methodology offers significant memory advantages over highly optimized, high-performance sparse direct solver libraries, while maintaining approximately comparable or slightly slower serial serial execution speed but with significantly better parallel and GPU processing prospects.

\end{abstract}
\vspace{10pt}
This work has been presented at the 2017 International Conference on Electromagnetics in Advanced Applications. The extension of this work has been presented at the 2019 IEEE International Symposium on Antennas and Propagation and USNC-URSI Radio Science Meeting \cite{moshfegh2019parallel}

\section{INTRODUCTION}
\label{sec:intro}

Finite Element Method (FEM) computational electromagnetic models result in very large but sparse linear systems that are often solved iteratively in lieu of direct factorization methods, \cite{direct-methods-for-sparse}, that use significantly more memory and have higher asymptotic computational complexity. Yet, when these models involve many independent excitations, or complex materials, or multi-scale or near resonance structures, even the most advanced preconditioned iterative methods can lose effectiveness and efficiency. This problematic behavior, coupled with the recent proliferation of computing cores and RAM memory has lead to a renewed interest in sparse matrix direct solution methods in certain FEM computations in CEM. For all other situations, sparse direct solvers remain indispensable, yet as a `hidden' component of preconditioned iterative methods e.g. corse-grid correction in multi-grid solvers.

Efficient implementations of classic sparse Cholesky or LDL$^T$ factorizations for symmetric systems, or sparse LU methods, make use of various data and temporal locality improving computational schemes such as the multi-frontal \cite{MUMPS1} or the left/right looking super-nodal \cite{Schenk2004} algorithms to reduce cache misses and significantly improve speed. When coupled with  advanced fill-in reducing matrix reordering schemes such as the multilevel nested dissection \cite{METIS} they can result in powerful solvers with impressive performance.   

This paper will explore {\em weather the conventional approach of using highly optimized high-performance sparse direct solver libraries is indeed the best available option for the direct solution of FEM problems?  }

\section{APPROACH}
\label{sec:appraoch}
In our attempt to answer this question, an alternative, domain decomposition (DD) based, direct solution strategy (with {\em numerical exact arithmetic}) for FEM models will be outlined and compared to state-of-the-art sparse matrix direct solvers. Unlike conventional approaches that are agnostic of the matrix physical model origins i.e. `black-box' solvers, this direct DD strategy relies on a judicious re-formulation of the  boundary value problem (BVP) as a decomposed BVP on a collection of domains. It is noted that the method is not restricted to Maxwell equations or electromagnetics, but is simply presented in the CEM context. The decomposed BVP and its subsequent mixed/hybrid variational formulation must be cast such that all  domains are always regular while maintaining the self-adjoint nature of the overall problem. This will prove critical in achieving an efficient direct solution. Subsequent discretization with suitably chosen FEM basis in terms of primal degrees-of-freedom (DoFs) in domain volumes and dual  DoF on domain interfaces leads to a large sparse arrowhead matrix that is then reduced in size by eliminating via many independent sparse direct factorizations the primal DoFs. This process is akin to the FETI reduction in iterative DDs, but with key appropriate modifications that will later facilitate a very efficient direct factorization. Formulation of the proposed approach can be found in \cite{moshfegh2016direct, moshfegh2017memory}.

The resultant auxiliary matrix is symmetric but orders of magnitude smaller than the original one, and is block-wise sparse (sparse collection of dense blocks), something computational very desirable in direct matrix factorizations because the high degree of data locality enables the use the blazing fast level 3 BLAS \cite{DONGARRA1990}. As such, this matrix is LDL$^T$ factorized via a very fast symbolic factorization step that is attributed clique graph of the block-wise sparse matrix, followed by a custom block LDL$^T$ with restricted Bunch-Kaufman pivoting \cite{BUNCH1977}.  Once the the auxiliary problem has been efficiently factorized, the solution of all or a subset of primal unknowns are recovered in an embarrassingly parallel manner via many smaller size sparse forward backward substitutions. 

\begin{figure}[h]
\begin{center}
\noindent
  \includegraphics[width=3.in]{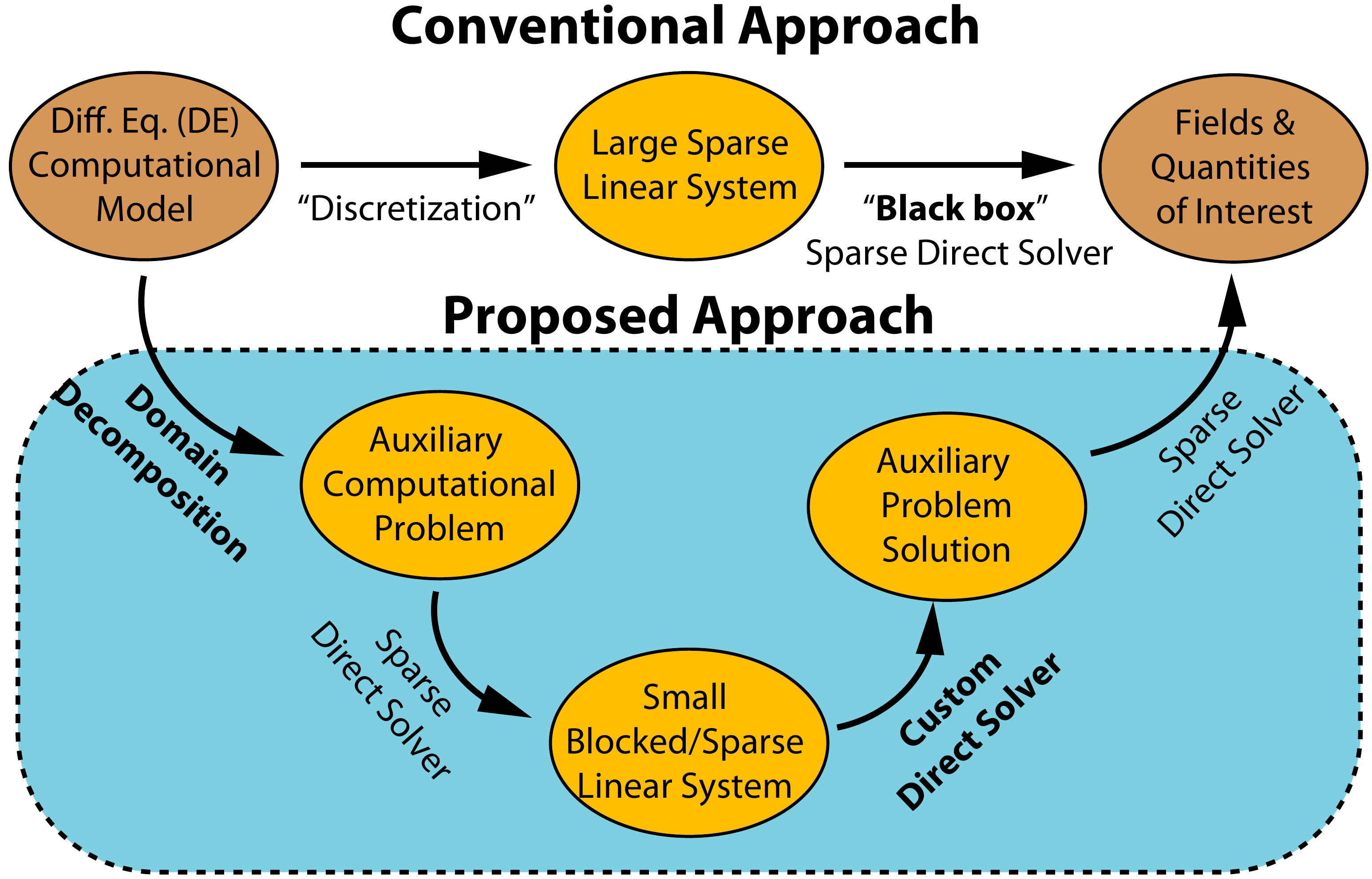}
\vspace{-10pt}
  \caption{Conceptual overview and comparison of conventional direct solutions of FEM models vs. the proposed direct solution approach.}\label{fig:approach}
\end{center}
\end{figure}

A conceptual overview of the proposed approach and its conventional counterpart are depicted in Fig. \ref{fig:approach}. Although the proposed  approach involves more computational steps, much like field computations with auxiliary potentials in electromagnetics, solving the auxiliary problem is  considerably easier than the conventional one-shot approach, but also attaining the auxiliary problem and recovering the full solution from it are relatively fast and embarrassingly parallel. In a sense out approach instead of striving to minimize fill-in to achieve better memory and speed, it attempts to maximize the concurrency in sparse matrix computations, as well as the total amount and order of subsequent dense matrix operations, thus promising very favorable parallel and GPU processing prospects.
%\vspace{-8pt}
\vspace{-20pt}

\section{COMPARISONS}
\label{sec:results}
\begin{figure}[h]
\begin{center}
  \includegraphics[width=3.in]{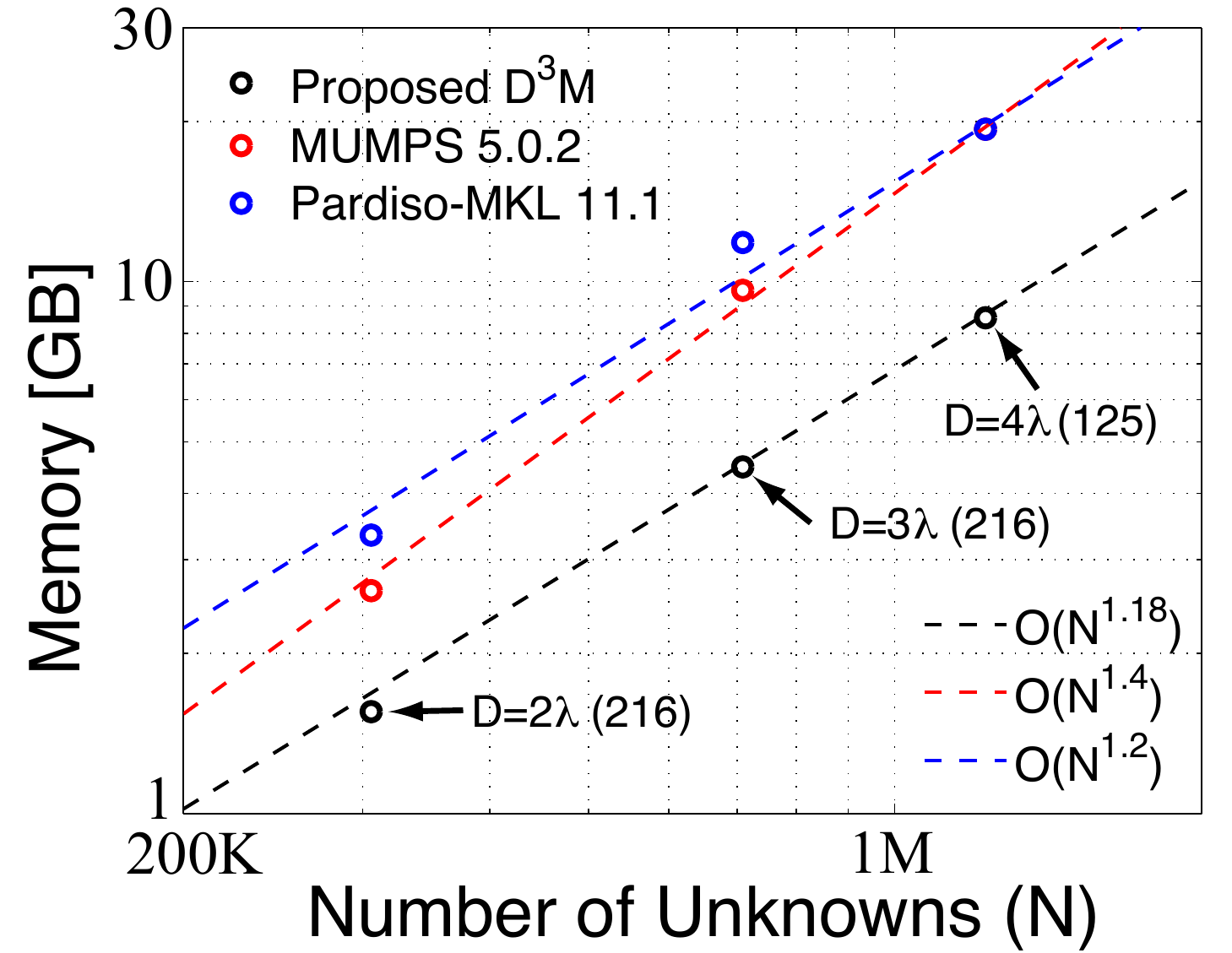} \\
  (a) \\
  \includegraphics[width=3.in]{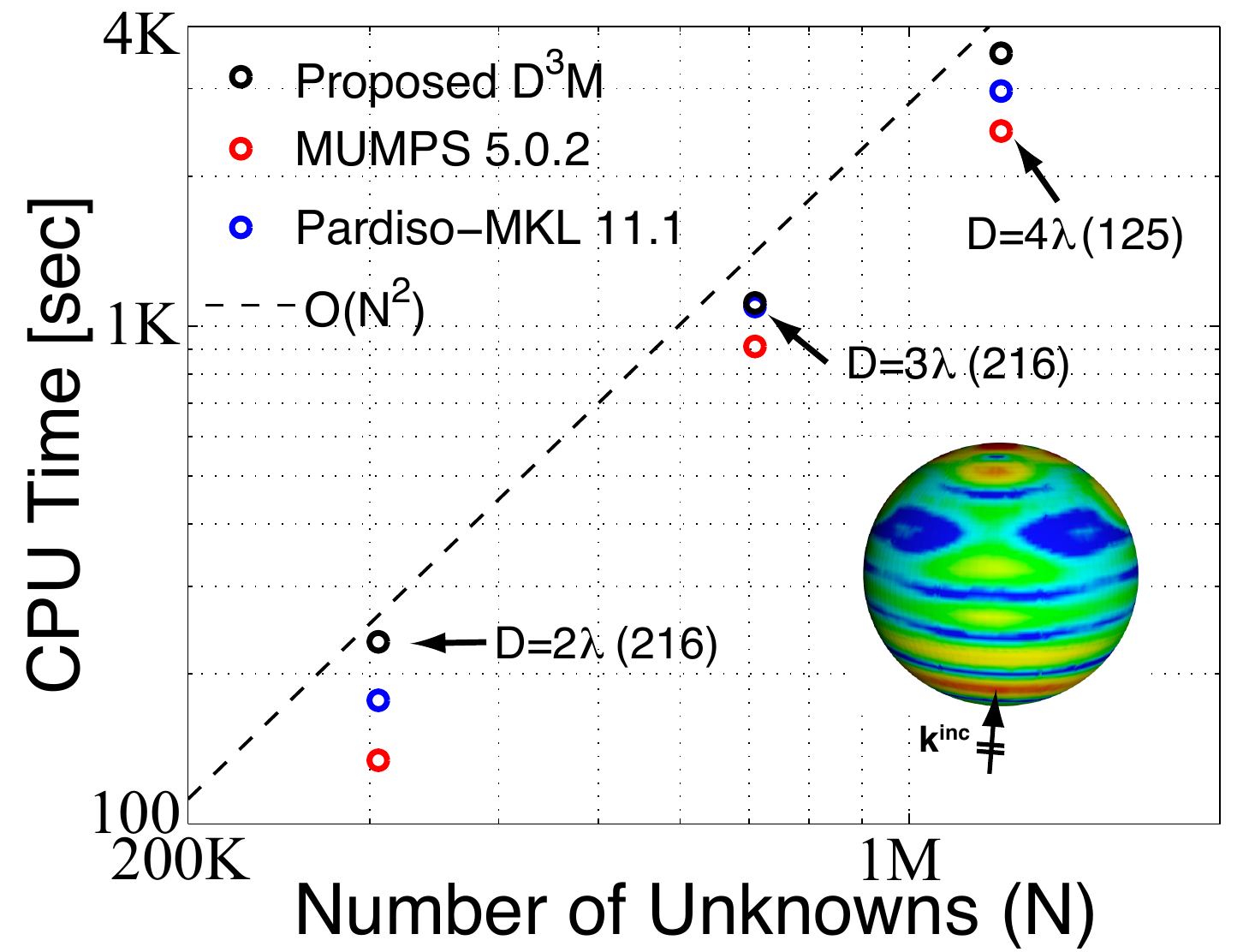}   \\
  (b)\\
    \includegraphics[width=3.in]{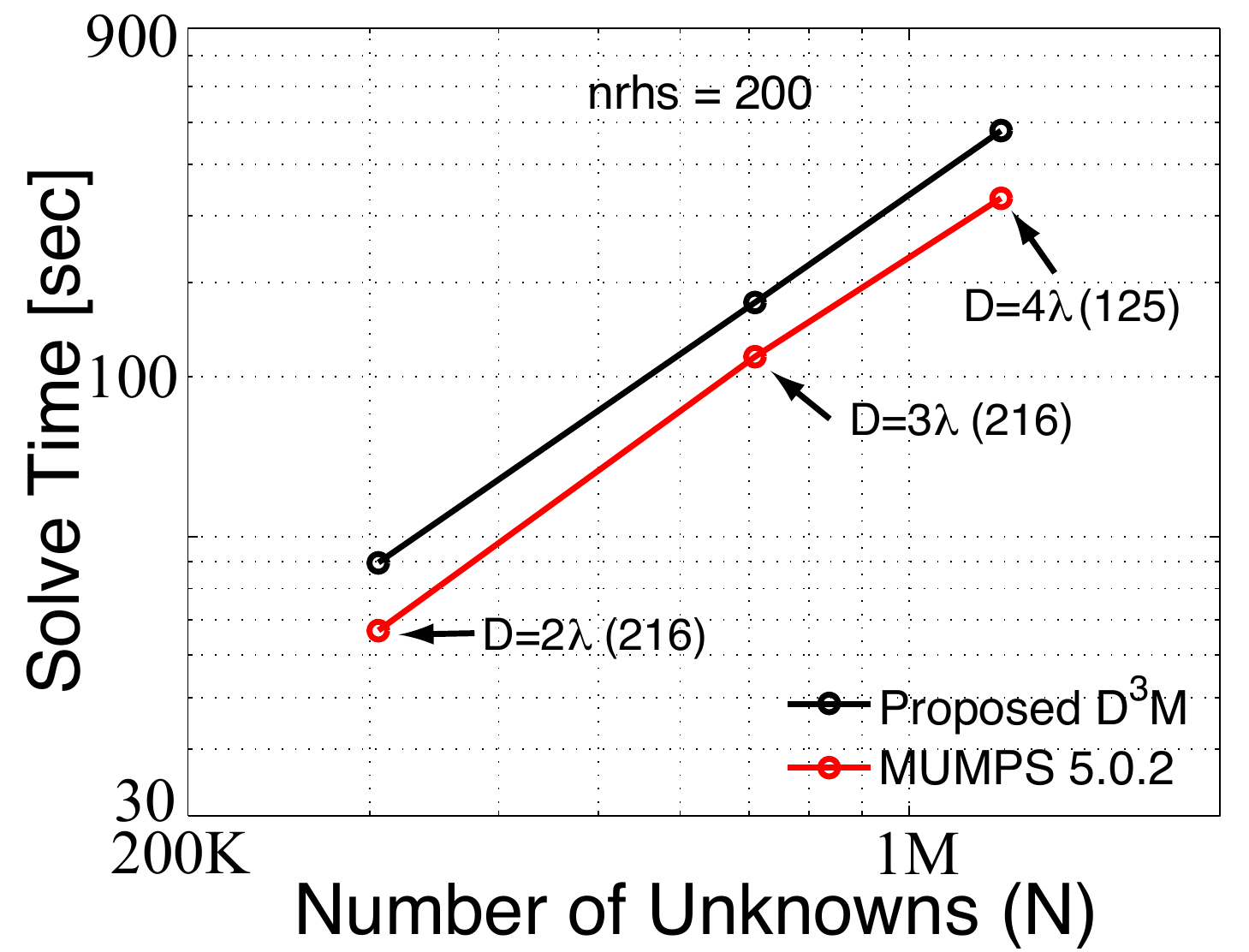}  \\
  (c)\\
\vspace{-10pt}
  \caption{Factorization memory and time comparison between the proposed direct DD approach and state-of-the-art direct solvers on progressively electrically larger dielectric spheres. (a) Memory; (b) CPU time (serial run); (c) forward/backward substitution time for 200 excitations (serial run). }\label{fig:sphere}
\end{center}
\end{figure}

\begin{figure}[h]
\begin{center}
  \includegraphics[width=3.in]{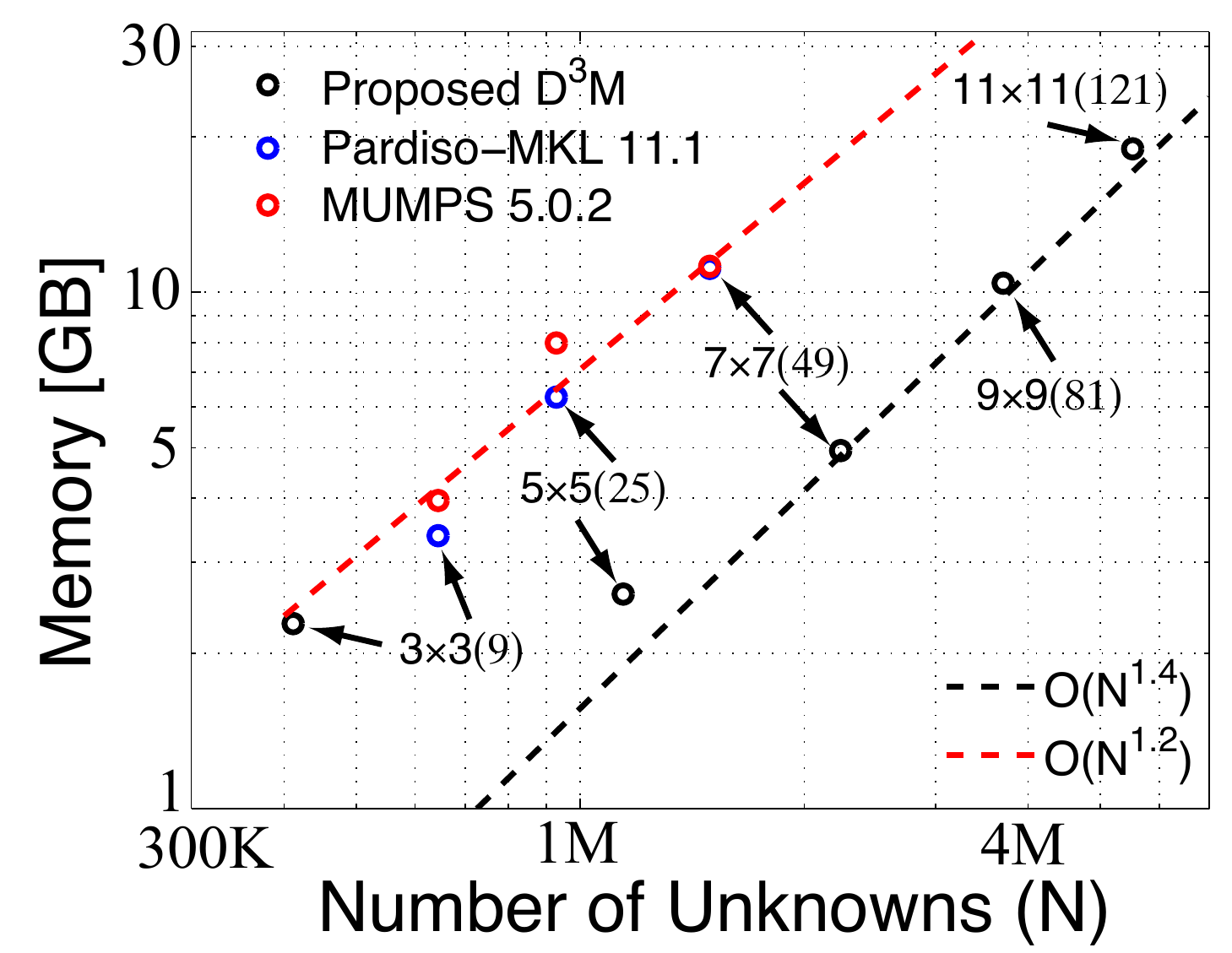}\\
  (a)\\
  \includegraphics[width=3.in]{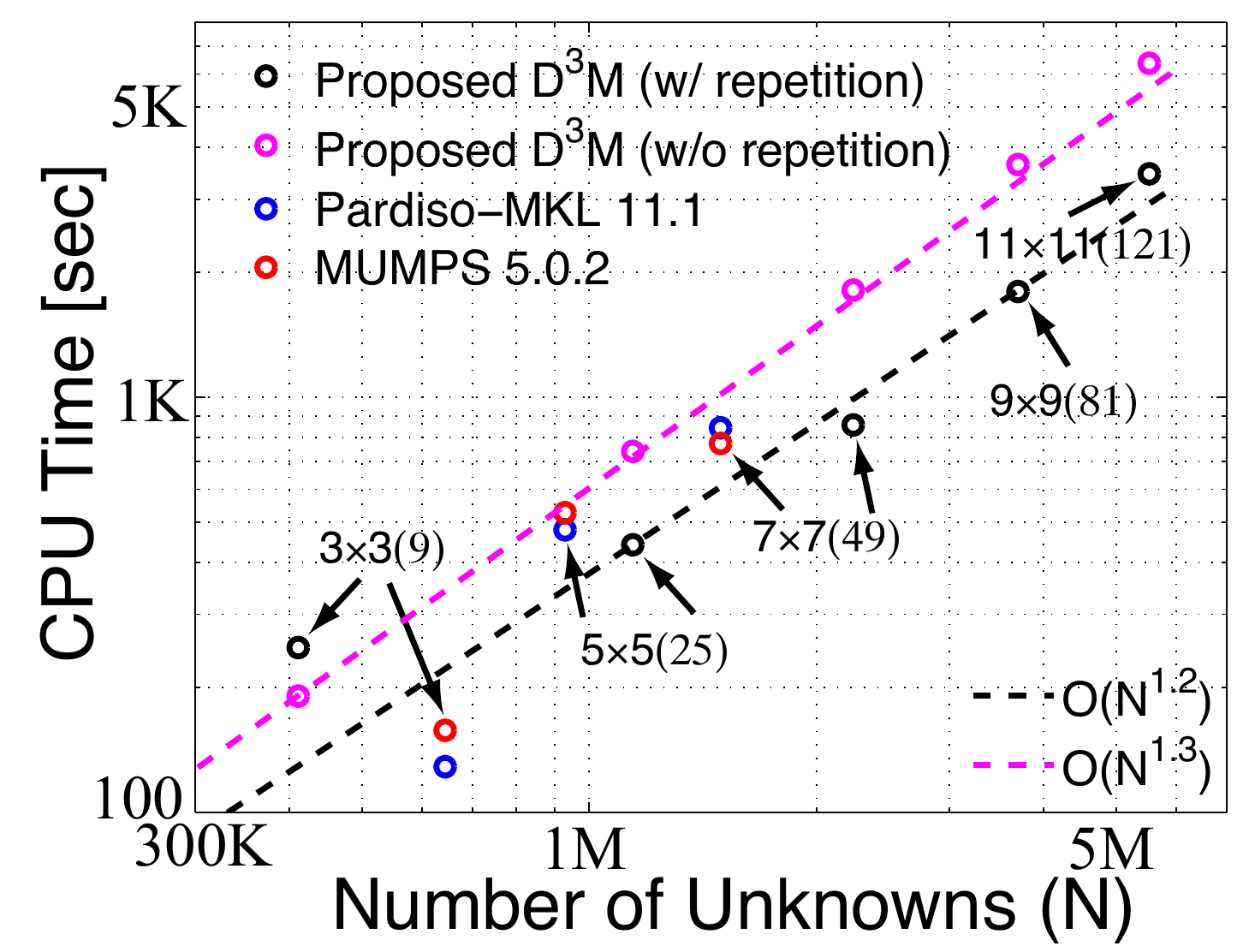}  \\
  (b)\\
    \includegraphics[width=3.in]{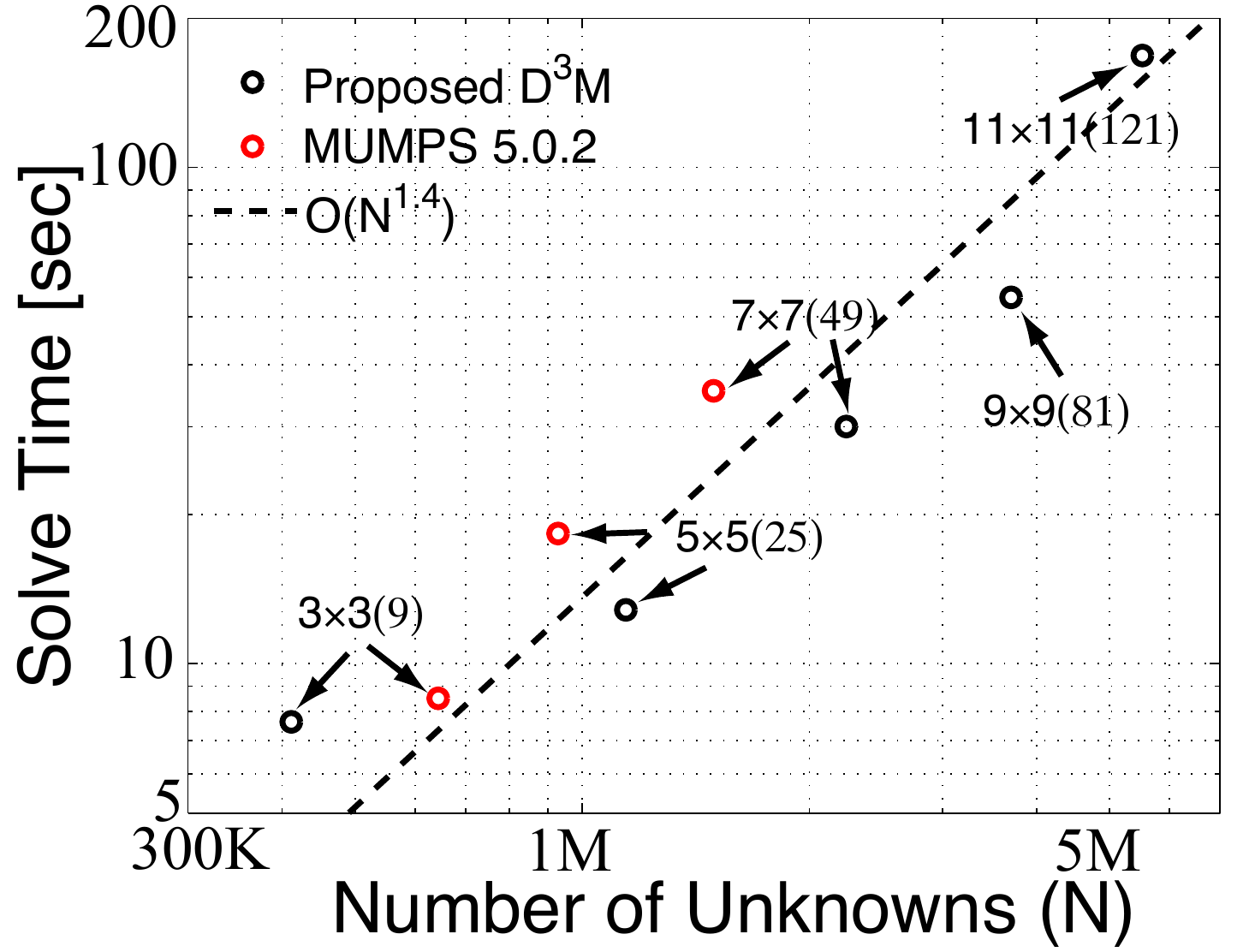}  \\
  (c)\\
\vspace{-10pt}
  \caption{Factorization memory and time comparison between the proposed direct DD approach and state-of-the-art direct solvers on progressively larger finite monopole antenna arrays. (a) Memory; (b) factorization time (serial run); (c) forward/backward substitution time (serial run). }\label{fig:monopole}
\end{center}
\end{figure}

Two low to moderate complexity unstructured FEM models are used as test cases to compare the proposed method with  MUMPS 5.0.2  \cite{MUMPS1}, and PARDISO-MKL11.1 \cite{MKL_PARDISO}  state-of-the-art `black-box' direct sparse matrix solvers with MeTiS 5.1.0 reordering for all methods. The first model involves a dielectric  sphere of progressively larger size, (domain increases in three dimensions), and a monopole antenna array problem again of progressively increasing size but also port number (excitations). The memory, factorization time and forward/backward substitution times of those test cases are presented in Figs \ref{fig:sphere} and \ref{fig:monopole}. In both cases, the purposed direct DD method shows a clear edge in memory requirements, which is a major improvement on a key direct method weakness  . Both factorization and forward backward substitution are on par for all methods with the proposed method being slightly slower.

%\vspace{-8pt}

 \section{CONCLUSIONS}
\label{sec:conclusions}
For the moderate size FEM problems examined here it was found that when memory is at premium, `black-box' direct solvers are not the best direct solution approach, the proposed direct DD method showed clear benefits as its memory appears to be comparable to preconditioned iterative solvers. When speed is important the highly optimized `back-box' libraries maintain a slight edge over the proposed direct DD and possibly iterative solvers. It is believe that direct DD solvers will have very favorable parallel and GPU processing prospects. 

%\vspace{-0.2in}
%\section*{Acknowledgments}
%I would like to thank my former and  current  Ph.D. students, Dr. G. Paraschos, Mr. W. Wang and Mr. J. Moshfegh for their contributions in this work.

%\begin{thebibliography}{9}
%\itemsep=0ex
%\bibitem{iceaa13} R.D.~Graglia, and G. Lombardi, ``Instructions for Paper Preparation and Submission to ICEAA-IEEE APWC'', electronic
%file available at \textsf{http://www.iceaa.net/iceaa-ieee\_apwc.pdf}, 2015.
%\bibitem{Latex2e} L.~Lamport, ``\LaTeX: a document preparation system'',
%Addison--Wesley Publishing Company, 2nd ed., 1994.
%\end{thebibliography}
\bibliographystyle{IEEEtran}
\bibliography{paper_bib.bib}

\begin{thebibliography}{10}
\providecommand{\url}[1]{#1}
\csname url@samestyle\endcsname
\providecommand{\newblock}{\relax}
\providecommand{\bibinfo}[2]{#2}
\providecommand{\BIBentrySTDinterwordspacing}{\spaceskip=0pt\relax}
\providecommand{\BIBentryALTinterwordstretchfactor}{4}
\providecommand{\BIBentryALTinterwordspacing}{\spaceskip=\fontdimen2\font plus
\BIBentryALTinterwordstretchfactor\fontdimen3\font minus
  \fontdimen4\font\relax}
\providecommand{\BIBforeignlanguage}[2]{{%
\expandafter\ifx\csname l@#1\endcsname\relax
\typeout{** WARNING: IEEEtran.bst: No hyphenation pattern has been}%
\typeout{** loaded for the language `#1'. Using the pattern for}%
\typeout{** the default language instead.}%
\else
\language=\csname l@#1\endcsname
\fi
#2}}
\providecommand{\BIBdecl}{\relax}
\BIBdecl

\bibitem{moshfegh2019parallel}
J.~Moshfegh, D.~G. Makris, and M.~N. Vouvakis, ``Parallel direct domain
  decomposition methods ({D3M}) for finite elements,'' in \emph{2019 IEEE
  International Symposium on Antennas and Propagation and USNC-URSI Radio
  Science Meeting}.\hskip 1em plus 0.5em minus 0.4em\relax IEEE, 2019, pp.
  777--778.

\bibitem{direct-methods-for-sparse}
I.~S. Duff, A.~M. Erisman, and J.~K. Reid, \emph{Direct Methods for Sparse
  Matrices}.\hskip 1em plus 0.5em minus 0.4em\relax New York, NY, USA: Oxford
  University Press, Inc., 1986.

\bibitem{MUMPS1}
P.~R. Amestoy, I.~S. Duff, J.~Koster, and J.-Y. L'Excellent, ``A fully
  asynchronous multifrontal solver using distributed dynamic scheduling,''
  \emph{SIAM Journal on Matrix Analysis and Applications}, vol.~23, no.~1, pp.
  15--41, 2001.

\bibitem{Schenk2004}
O.~Schenk and K.~Gartner, ``Solving unsymmetric sparse systems of linear
  equations with {PARDISO},'' \emph{Future Generation Computer Systems},
  vol.~20, no.~3, pp. 475 -- 487, 2004.

\bibitem{METIS}
G.~Karypis and V.~Kumar, ``A fast and highly quality multilevel scheme for
  partitioning irregular graphs,'' \emph{IAM Journal on Scientific Computing},
  vol.~20, no.~1, pp. 359--392, 1999.

\bibitem{moshfegh2016direct}
J.~Moshfegh and M.~N. Vouvakis, ``Direct domain decomposition method ({D3M})
  for finite element electromagnetic computations,'' in \emph{Antennas and
  Propagation (APSURSI), 2016 IEEE International Symposium on}.\hskip 1em plus
  0.5em minus 0.4em\relax IEEE, 2016, pp. 1127--1128.

\bibitem{moshfegh2017memory}
------, ``A memory-efficient sparse direct solver with applications in {CEM},''
  in \emph{2017 IEEE International Symposium on Antennas and Propagation \&
  USNC/URSI National Radio Science Meeting}.\hskip 1em plus 0.5em minus
  0.4em\relax IEEE, 2017, pp. 1577--1578.

\bibitem{DONGARRA1990}
J.~J. Dongarra, J.~Du~Croz, S.~Hammarling, and I.~S. Duff, ``A set of level 3
  basic linear algebra subprograms,'' \emph{ACM Trans. Math. Softw.}, vol.~16,
  no.~1, pp. 1--17, Mar. 1990.

\bibitem{BUNCH1977}
J.~R. Bunch and L.~Kaufman, ``Some stable methods for calculating inertia and
  solving symmetric linear systems,'' \emph{Math. Comp.}, vol.~31, no. 137, pp.
  163--179--41, 1977.

\bibitem{MKL_PARDISO}
Intel, ``Intel math kernel library ver. 11.1,'' Intel Corporation, Santa Clara,
  CA, USA, 2009.

\end{thebibliography}

\end{document}